\documentclass[%
 reprint,
 superscriptaddress,
 amsmath,amssymb,
 aps,
 prx,
]{revtex4-2}

\usepackage{graphicx}

\usepackage{booktabs}
\usepackage{bm}
\usepackage{color}
\usepackage[normalem]{ulem}

\begin{document}

\title{Analytical Fock-State Generation and $\mathrm{SWAP}$ using a Rabi-Driven Transmon}
\author{N. Karaev}
\email[Contact author: ]{natan.karaev@campus.technion.ac.il}
\author{E. Blumenthal}
\author{S. Hacohen-Gourgy} 
\affiliation{Department of Physics, Technion - Israel Institute of Technology, Haifa 32000, Israel}

\begin{abstract}
Deterministic Fock-state generation and inter-mode SWAP are foundational primitives for bosonic quantum computing, yet most implementations rely on numerically optimized pulses, per-state calibration, strong dispersive coupling, or higher transmon levels, each adding control overhead that grows with system size. We present an analytical, calibration-light protocol operating entirely within the two-level $g$--$e$ manifold of a weakly dispersively coupled transmon. A Rabi drive on the qubit, combined with a single sideband tone per mode, synthesizes an on-demand Jaynes-Cummings interaction whose entire family of pulse times follows the closed-form scaling $\tau_n=\tau_1/\sqrt{n}$. Once the single base time $\tau_1$ is set, every higher-$n$ operation is fixed analytically, with no per-state retuning, shelving, or numerical optimization. On a superconducting flute cavity with two high-$Q$ modes, we deterministically prepare Fock states through $|n{=}5\rangle$, realize an inter-mode SWAP characterized on vacuum, single-photon, and coherent-state inputs, and generate and coherently swap the dual-rail Bell state $(|1,0\rangle+|0,1\rangle)/\sqrt{2}$, confirming that the operation preserves inter-mode coherence. Because the pulses are constant-amplitude and free of per-state optimization, the achievable fidelity is set directly by ancilla coherence and drive-ramp duration; a master-equation analysis isolates these hardware factors and shows that the analytical scaling itself imposes no obstacle to high-fidelity operation at high $n$. Requiring only one sideband line per mode and a single Rabi drive, the protocol is well suited to weakly coupled, high-$Q$ 3D architectures where calibration economy and analytical pulse design are at a premium.
\end{abstract}

\maketitle


\section{\label{sec:intro}Introduction}

Fock states are a foundational computational resource for bosonic quantum
computing~\cite{krastanov_universal_2015,michael_new_2016}, and the
deterministic preparation and inter-mode transfer of these states underpin
dual-rail encodings, error-correctable bosonic
codes, and modular architectures based on cavity-to-cavity state
transfer~\cite{teoh_dual-rail_2023,gottesman_demonstrating_1999,divincenzo_physical_2000,
axline_-demand_2017,zhou_quantum_2024}. Achieving these primitives reliably on
modes that are only weakly coupled to their ancilla, a regime favored for
preserving cavity coherence, remains a central challenge.

A range of high-performing protocols has emerged to address this challenge,
each with a characteristic calibration cost. Numerically optimized pulses
(GRAPE, ECD-control) achieve high fidelities but require per-state
optimization on a system-specific
Hamiltonian~\cite{eickbusch_fast_2022,kudra_experimental_2025}. SNAP-based
schemes rely on photon-number-selective qubit rotations and dedicated
calibration of each selective pulse~\cite{landgraf_fast_2023}. Charge-driven
$g$--$f$ sideband protocols exploit the third transmon level to attain fast,
multimode control and reach high binomial-code-state fidelities~\cite{huang_fast_2026}, at the cost of calibrating multiple
sideband transitions, shelving operations, and photon-number-selective drives.
Across these approaches, raw fidelity has improved markedly, while the number
of independently tuned parameters per state has not.

In this work we explore a complementary point in this design space: a protocol
in which a single closed-form scaling law replaces per-state calibration. A
Rabi drive on a weakly coupled transmon, combined with a sideband tone
detuned by the Rabi frequency, induces a controllable Jaynes-Cummings interaction~\cite{murch_cavity-assisted_2012,
wang_observation_2022,kitzman_phononic_2022,lu_multipartite_2022}. The
resulting pulse times satisfy
an analytical relation, and thus generating Fock states up to
arbitrary $n$ requires only a single time calibration. The same mechanism,
applied across two modes, realizes an inter-mode SWAP whose dynamics we
analyze. We demonstrate deterministic preparation of
$|n{=}1\rangle$ through $|n{=}5\rangle$, an inter-mode SWAP characterized on vacuum, single-photon, and
coherent-state inputs, and the generation and swapping of a
dual-rail Bell state. We further quantify, both experimentally and through master-equation
simulation, the regimes to which this analytical, calibration-light approach is best
suited.

The structure of this paper is as follows. Sec.~\ref{sec:theory} develops
the theoretical framework. Sec.~\ref{sec:landscape} situates our approach
within the existing protocol landscape and identifies the regimes in which
each approach is most natural. Sec.~\ref{sec:results} reports experimental
generation, SWAP, and Bell-state results, with quantitative comparison to the
ideal-system performance derived from simulation. Sec.~\ref{sec:limits}
discusses practical limits and the path to competitive fidelities. Detailed
derivations and supplementary simulations are deferred to the appendices.

\section{Theory}\label{sec:theory}

\begin{figure}
    \centering
    \includegraphics[width=1\linewidth]{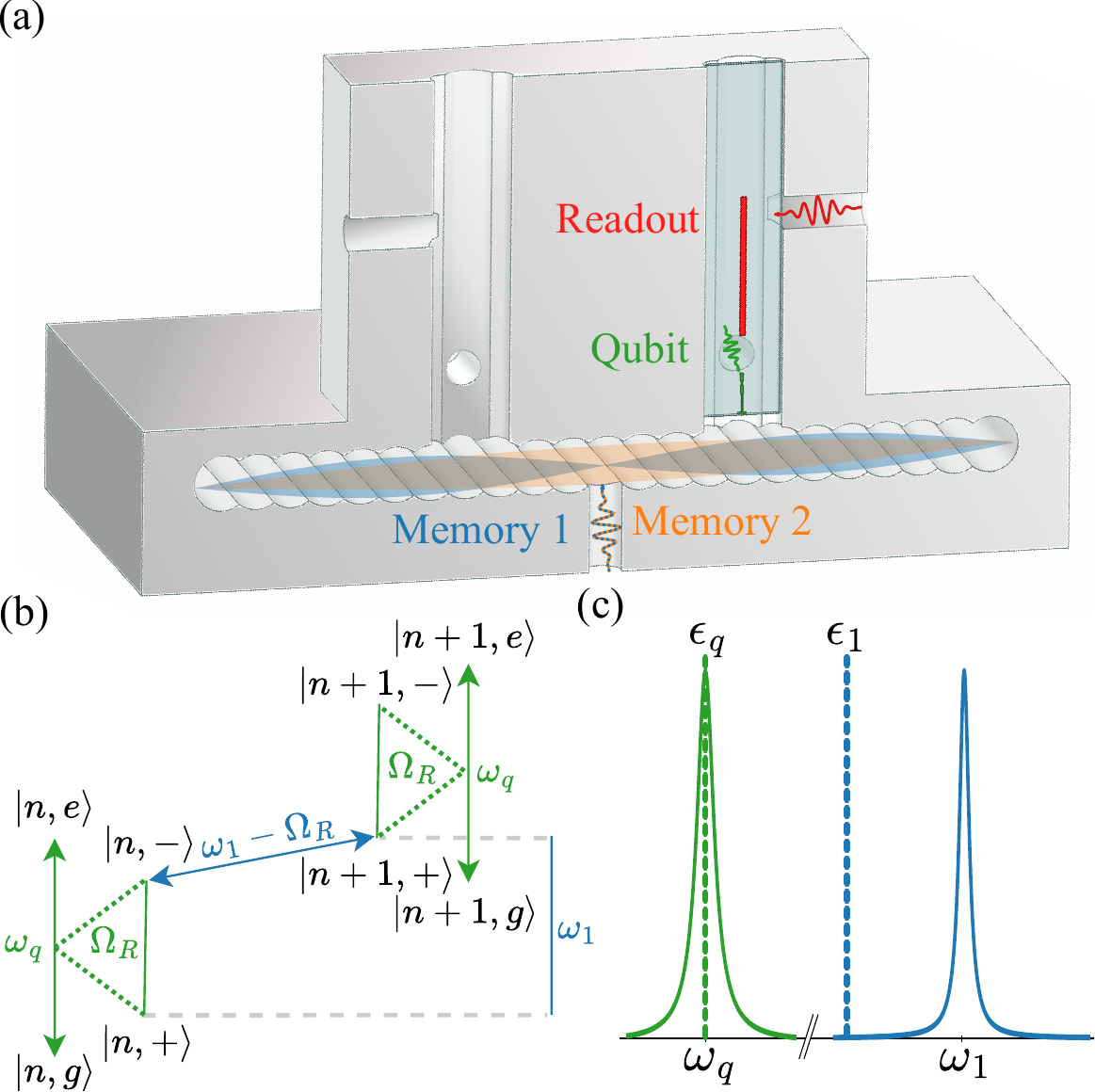}
    \caption{System diagrams. The elements representing the Readout, Qubit, Memory 1, and Memory 2 components are red, green, blue, and orange, respectively. (a) Cross section view of the aluminum flute cavity used in the measurements. A sapphire chip contains a stripline readout resonator and a qubit. The Memory 1 and Memory 2 are The $\mathrm{TE}_{102}$ and $\mathrm{TE}_{101}$ modes inside the flute cavity, respectively. (b) Energy level diagram showing the coupling of Memory 1 and the qubit. Solid bidirectional lines represent drives. The frequency of each drive is noted. Undirectional solid lines represent energy level differences. Dotted lines represent the emergence of the dressed states due to the Rabi drive. The dashed gray lines are a guide to the eye. Memory 2 and the Readout are coupled similarly. (c) Frequency domain mapping of the coupling of Memory 1 and the qubit. Solid lines represent the frequency responses of the components. Dashed lines represent drives. Memory 2 and the Readout are driven similarly.}
    \label{fig:system}
\end{figure}


\begin{figure*}
    \centering
    \includegraphics[width=1\linewidth, trim = 10 0 0 0, clip]{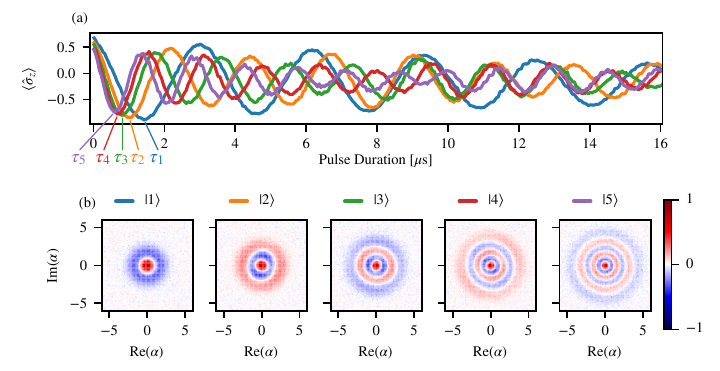}
    \caption{Calibration and measurement of generated Fock states. (a) Optimization of the duration of the different $\tau_n$ pulse times performed by measuring $\left<\hat{\sigma}_z\right>$ and identifying the minimum value. This measurement is done separately for each $\tau_n$ of each Fock state. $\tau_1$, $\tau_2$, $\tau_3$, $\tau_4$, and $\tau_5$ are $1.42\mu\mathrm{s}$, $1.01\mu\mathrm{s}$, $0.81\mu\mathrm{s}$, $0.69\mu\mathrm{s}$, and $0.61\mu\mathrm{s}$ respectively. (b) Wigner characteristic function measurements for generated Fock states. The fidelity values are $91.64\%$, $82.38\%$, $75.82\%$, $69.39\%$, and $62.98\%$, for $\left|1\right>$,$\left|2\right>$,$\left|3\right>$,$\left|4\right>$ and $\left|5\right>$, respectively.}
    \label{fig:GeneratedFocks}
\end{figure*}

\begin{figure}
    \centering
    \includegraphics[width=1\linewidth, trim = 58 5 35 0, clip]{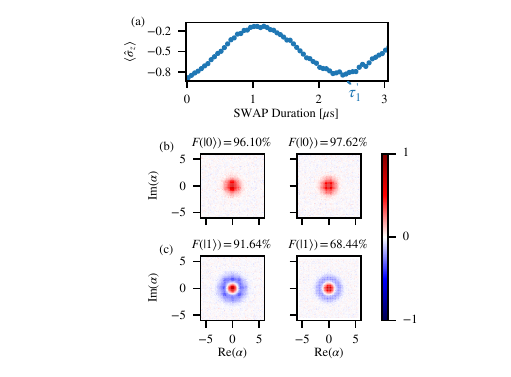}
    \caption{Inter-mode SWAP characterization.
    (a) Optimization of the SWAP duration $\tau'_1$ via measurement of
    $\langle\hat{\sigma}_z\rangle$; the extremum is at $\tau'_1 = 2.27~\mu$s,
    in agreement with the predicted $\sqrt{2}\,\tau_1$. (b) Wigner
    characteristic functions of Memory~1 (left) and Memory~2 (right) following
    SWAP with both modes initialized in vacuum, with fidelities to $|0\rangle$
    of $96.10\%$ and $97.62\%$. The pre-SWAP initial fidelities to vacuum of Memory 1 and Memory 2 are $98.04\%$, and $99.51\%$, respectively. (c) Wigner characteristic function of Memory~1
    before (left, $F(|1\rangle)=91.64\%$) and Memory~2 after (right,
    $F(|1\rangle)=68.44\%$) SWAP of a prepared $|1\rangle$ state. Under the
    assumption of multiplicatively independent preparation and SWAP errors,
    the latter corresponds to a SWAP-step state-transfer fidelity of
    $\sim\!74.68\%$; this single-input estimate does not constitute an average
    gate fidelity (see Sec.~\ref{sec:resFST}).}
    \label{fig:FST}
\end{figure}

\begin{figure}
    \centering
    \includegraphics[width=1\linewidth, trim = 30 0 20 0, clip]{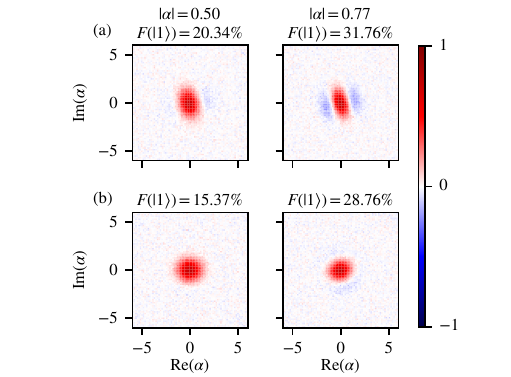}
    \caption{SWAP on small coherent states, tracking transfer of the
    $|1\rangle$ component. (a) Two coherent states ($|\alpha|=0.50$ and
    $|\alpha|=0.77$) initialized in Memory~1, dominated by their $|0\rangle$
    and $|1\rangle$ components; indicated fidelities are to the $|1\rangle$
    component. (b) Memory~2 following SWAP, showing transfer of the
    $|1\rangle$ population. The output is not a simple displaced state owing
    to the $n$-dependent SWAP dynamics of the higher Fock components present
    in a coherent input; see Sec.~\ref{sec:resFST}.}
    \label{fig:SwapCoherent}
\end{figure}

\begin{figure*}
    \centering
    \includegraphics[width=1\linewidth, trim = 7 0 10 0, clip]{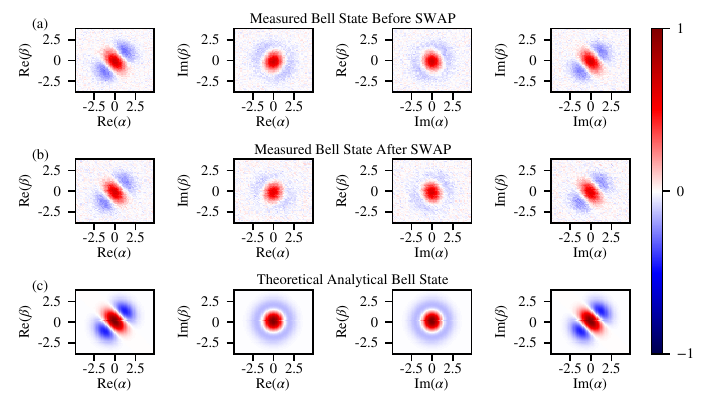}
    \caption{Joint Wigner characteristic Function of the $\left(\left|1,0\right>+\left|0,1\right>\right)/\sqrt{2}$ Bell state. The axis of the displacement is noted by the label of each axis. $\alpha$ and $\beta$ represent conditional displacement amplitudes along different axes on Memory 1 and 2, respectively.(a) Measured joint Wigner characteristic function after Bell state generation. (b) Measured joint Wigner characteristic function after Bell state generation and SWAP.  (c) Theoretical analytical plot for the joint Wigner characteristic function of the $\left(\left|1,0\right>+\left|0,1\right>\right)/\sqrt{2}$ Bell state, computed in closed-form without decoherence.}
    \label{fig:JointChar}
\end{figure*}

\begin{figure*}
    \centering
    \includegraphics[width=0.8\linewidth,trim = 60 10 80 10, clip]{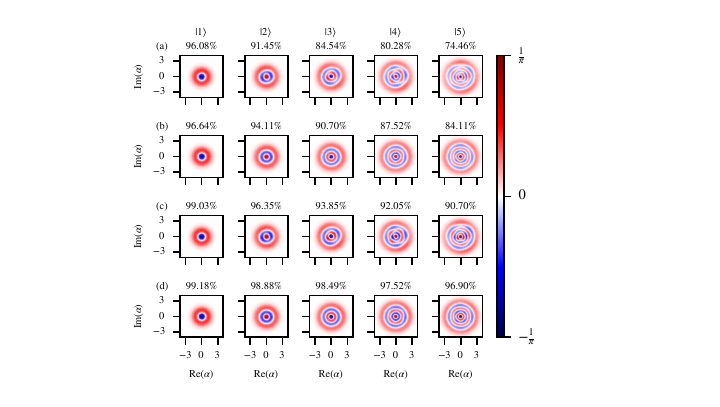}
    \caption{Ideal simulation with decoherence of Fock State Generation with different ramp lengths and decoherence times. (a) Original system parameters $T_{1_q} = 22.8~\mu\mathrm{s}$, $T_{2_q}^{\mathrm{echo}}   = 21.8~\mu\mathrm{s}$, $T_{1_1} = 145~\mu\mathrm{s}$. Ramp lengths are $200~\mathrm{ns}$. (b) $T_{1_q} = 22.8~\mu\mathrm{s}$, $T_{2_q}^{\mathrm{echo}}   = 21.8~\mu\mathrm{s}$, $T_{1_1} = 145~\mu\mathrm{s}$. Ramp lengths are $20~\mathrm{ns}$. (c) $T_{1_q} = 228~\mu\mathrm{s}$, $T_{2_q}^{\mathrm{echo}}   = 218~\mu\mathrm{s}$, $T_{1_1} = 1450~\mu\mathrm{s}$. Ramp lengths are $200~\mathrm{ns}$. (d) $T_{1_q} = 228~\mu\mathrm{s}$, $T_{2_q}^{\mathrm{echo}}   = 218~\mu\mathrm{s}$, $T_{1_1} = 1450~\mu\mathrm{s}$. Ramp lengths are $20~\mathrm{ns}$.}
    \label{fig:simFockGen}
\end{figure*}

Fock state generation and transfer are performed on a system shown in Fig.~\ref{fig:system}(a). The system consists of two long-lived flute cavity~\cite{chakram_seamless_2021} modes termed Memory 1 and Memory 2, a qubit, and a short-lived stripline resonator harmonic mode utilized for readout and reset operations termed Readout. The coupling scheme used in this work is an evolution of the high-\textit{Q} cavity mode reset method shown in~\cite{karaev_cavity-mode_2025,blumenthal_experimental_2026}, itself an evolution of the cooling method shown in~\cite{murch_cavity-assisted_2012}. In these works a Rabi-driven qubit is coupled to different harmonic modes via sideband drives detuned by the Rabi frequency. This type of coupling allows fast at-will transfer of excitations in systems with weak coupling. Let us construct the system Hamiltonian from the frame of reference of the sideband drive of each of the harmonic modes and from the Rabi drive frame of reference for the qubit,

\begin{equation}
    \begin{split}
        H=-\Omega_R\hat{a}_r^\dagger \hat{a}_r-\Omega_R\hat{a}_1^\dagger \hat{a}_1-\Omega_R\hat{a}_2^\dagger \hat{a}_2
        -\frac{\Omega_R}{2}\hat{\sigma}_x\\-\chi_r \hat{a}_r^\dagger \hat{a}_r\hat{\sigma}_z-\chi_1 \hat{a}_1^\dagger \hat{a}_1\hat{\sigma}_z-\chi_2 \hat{a}_2^\dagger \hat{a}_2\hat{\sigma}_z\\+\epsilon_r(\hat{a}_r^\dagger+\hat{a}_r)+\epsilon_1(\hat{a}_1^\dagger+\hat{a}_1)+\epsilon_2(\hat{a}_2^\dagger+\hat{a}_2),
    \end{split}
    \label{eq:SystemHamilt}
\end{equation}

where $\Omega_R$ is the Rabi frequency, $a$ are the harmonic mode annihilation operators, $\chi$ is the dispersive shift coefficient, $\epsilon$ is the drive power, and the subscripts $r,1,2$ refer to the Readout mode, Memory 1 mode, and Memory 2 mode, respectively.

When drive is applied at a constant power, the coherent state amplitudes of the steady states of the different harmonic modes are
\begin{equation}
        \bar{a}_{1,2}=\frac{\epsilon_{1,2}}{\Omega_R}~, ~\bar{a}_r=\frac{\epsilon_r}{\Omega_R+\frac{i\kappa_r}{2}},
    \label{eq:CavOccupation}
\end{equation}
where $\kappa$ is the Readout linewidth. We assume that the memory modes' linewidths are negligible compared to $\Omega_R$. We can now shift the annihilation operators by their steady state. The new dressed states of the Rabi-driven qubit are now $\left|+\right>$ and $\left|-\right>$ states. Then, after applying the rotating wave approximation to discard terms rotating at $\Omega_R$ and shifting to the Hadamard frame, we are left with

\begin{equation}
    \begin{split}
        H=\sum_{i=r,1,2}^{}{(g_i^\star \hat{a}_i\hat{\sigma}_++g_i\hat{a}_i^\dagger\hat{\sigma}_-)},
        \end{split}
    \label{eq:SimpleHamilt}
\end{equation}

where $g_i\equiv|\chi_i|\bar{a}_i$ is the coupling coefficient between each mode and the qubit, for the $i$ subscripts $r$, $1$, and $2$. From this form of the Hamiltonian it is possible to see that this type of setup allows for at-will Jaynes-Cummings excitation exchange between the qubit and each harmonic mode, as shown schematically in Fig.~\ref{fig:system}(b). While the coupling rate is theoretically limited only by drive power and the limits of the dispersive regime, in practice the rotating wave approximation requires $g\ll\Omega_R$, and the maximum $\Omega_R$ is limited by the maximum qubit drive of the system. Additionally, strong drives can couple the system's components to Two-Level Systems (TLSs) or produce higher order effects. Thus, a system tailored to avoid such transitions could have very short operation times, even for comparatively small $\chi$ values. The frequency responses and drives for a single coupling term are shown in Fig.~\ref{fig:system}(c).

\subsection{Fock State Generation Theory}\label{sec:FSGTheory}
The qubit and the relevant memory mode are initialized in the ground and vacuum states, respectively. The appropriate sideband drive for coupling is ramped up, and subsequently we excite the qubit to the higher-energy dressed state. The Rabi drive is then activated. For the duration of the Rabi drive, excitation exchange is in effect. If this duration is approximately $\tau_1=\pi/2g$, the single excitation contained in the dressed states of the qubit is transferred to the memory mode. Thus, the memory mode contains the Fock state $\left|n=1\right>$. It is possible to use a similar logic to generate higher photon count Fock states: Since the dressed states are the $\left|+\right>$ and $\left|-\right>$ states, we can inject the qubit with an excitation by performing a virtual $Z$ rotation of $\pi$, i.e. by flipping the sign of the Rabi drive. Activating the Rabi drive for a new time period $\tau_2$, the transition that
populates $\left|n\right>$ couples at rate $g\sqrt{n}$ in
Eq.~\ref{eq:SimpleHamilt}, so that $\tau_n=\tau_1/\sqrt{n}$.
After time $\tau_2$, the memory mode now contains $\left|n=2\right>$. This process can be repeated as necessary in order to create higher photon count Fock states, with Rabi drive times shortened by a factor of $\sqrt{n}$ compared to $\tau_1$ for each increase in the photon count. The process ends with a $\pi/2$ pulse in phase with the final Rabi drive to shift the qubit from the lower-energy dressed state back to its lab frame ground state.

\subsection{Fock State SWAP Theory}\label{sec:theoFST}
In order to perform Fock state SWAP we first activate the sideband drives for both memory modes, such that $g\equiv g_1=g_2$, followed by a Rabi drive. Initializing one memory mode at some Fock state and the other memory mode at vacuum, while initializing the qubit at the lower-energy dressed state, the dynamics of the system will evolve as an antisymmetric mode. Thus, after some time period $\tau'_n$ for initial state $\left|n\right>$ the memory modes' states will have swapped. In a single-excitation $\left|n=1\right>$ case, the dynamics reduce to coupling at rate $\sqrt{2}g$, yielding $\tau'_1=\sqrt{2}\tau_1$. Because each excitation exchange term in Eq.~\ref{eq:SimpleHamilt} adds a phase of $\pi/2$, the swapped states have a total phase of $\pi$, thus the operation performed is a $-\mathrm{SWAP}$ gate, which is a SWAP gate up to a virtual rotation on one of the modes (see Appendix~\ref{app:BDM}). 
For Fock states with $n>1$, the system dynamics no longer reduce to
linear excitation exchange. Because the qubit can hold at most a single
excitation, the system effectively couples two harmonic oscillators
through a nonlinear ancilla, and the time $\tau'_n$ to perform a SWAP
is not given by a simple analytical factor as in
Sec.~\ref{sec:FSGTheory}. There will also always be a degree of non-ideality in the transferred state, leading to a potential decrease in the fidelity. The dynamics of the Hamiltonian in Eq.~\ref{eq:SimpleHamilt} mean that the effective phase of such an interaction will be $\pi$ and $0$ for odd and even $n$, respectively. This means that for odd and even photon numbers this interaction will be a $\mathrm{SWAP}$ gate and a $-\mathrm{SWAP}$ gate, respectively. These gates are equivalent up to single-mode phase.

This interaction can also be leveraged to create a Bell state. If the qubit is initialized in the higher-energy dressed state and the SWAP pulses are performed for half the time $\tau'_1$,  the total state becomes a maximally entangled $\left(\left|1,0\right>+\left|0,1\right>\right)/\sqrt{2}$ Bell state. 

 Simulations from which the SWAP times can be calculated in some $n\geq2$ cases are shown in Appendix~\ref{app:simFST}. The full dynamics of the SWAP gate and Bell state generation are theoretically derived in Appendix~\ref{app:BDM}.

\section{Protocol Landscape and Comparison}\label{sec:landscape}

Recent years have seen rapid progress in deterministic Fock- and bosonic-code
state preparation in circuit QED. To clarify the niche occupied by the present
work, we summarize four representative classes of protocol, together with our
own, in Table~\ref{tab:landscape}. The classes differ along several
practically important axes: the ancilla manifold employed, the dispersive
coupling regime, the structure of the control pulses, and the calibration
cost per additional state.

\begin{table}[t]
\renewcommand{\arraystretch}{1.5}
\setlength{\tabcolsep}{4pt}
\caption{\label{tab:landscape}%
Broader landscape of deterministic Fock-state and bosonic-code
preparation protocols. Figures are illustrative of typical reported
implementations and intended for methodological orientation,
not strict ranking.}
\begin{tabular}{p{1.7cm}p{1.5cm}p{2.1cm}p{1.6cm}}
\hline\hline
\textbf{Protocol class} & \textbf{Ancilla manifold} & \textbf{Per-state calibration} & \textbf{Typical regime} \\
\hline
This work &
 $g$--$e$ &
 None ($\tau_n\!=\!\tau_1/\sqrt{n}$) &
 Weak $\chi$ \\[2pt]
$g$--$f$ sideband~\cite{huang_fast_2026} &
 $g$--$f$ + shelving &
 Per-transition tuning &
 Weak $\chi$ \\[2pt]
ECD-control~\cite{eickbusch_fast_2022} &
 $g$--$e$ &
 Re-optimize $\{\beta_k\}$ &
 Weak $\chi$ \\[2pt]
Optimal control~\cite{kudra_experimental_2025} &
 $g$--$e$ &
 Re-run numerical opt. &
 Moderate to strong $\chi$ \\[2pt]
SNAP + disp.~\cite{landgraf_fast_2023} &
 $g$--$e$ &
 New selective pulses &
 Strong $\chi$ \\
\hline\hline
\end{tabular}
\end{table}

Two structural advantages emerge from Table~\ref{tab:landscape}. First, our protocol is the only one that simultaneously operates purely in the $g$--$e$ manifold, utilizes weak static $\chi$, and relies on a single analytical scaling law rather than per-state numerical optimization. Second, because it removes complex pulse shaping, the protocol's performance is strictly bounded by hardware coherence rather than control complexity. While protocols employing numerical optimization or higher transmon levels can
reach higher absolute fidelities, our simulations (Sec.~\ref{sec:limits}) show
that the gap to those demonstrations is set by device-specific ancilla
coherence and drive ramps, not by the underlying methodology.

Thus, this protocol is best suited to (i) systems with
small static $\chi$ (tens of kHz), where strong-coupling protocols become
slow or impractical; (ii) systems whose ancilla $|f\rangle$ level is short-lived,
charge-sensitive, or spectrally inconvenient, ruling out $g$--$f$ sidebands;
(iii) dual-rail or multi-mode experiments in which only a small handful of
Fock-state operations per mode are required, and the overhead of
state-dependent pulse re-calibration becomes the dominant control cost;
(iv) experiments in which an analytical scaling law is required for
extrapolation, calibration robustness, or theoretical analysis. The protocol
also requires only a single sideband line per mode and a single Rabi drive
on the qubit, simplifying control electronics.

The experimental demonstrations of Sec.~\ref{sec:results} are situated within
the favorable regime; the limits we observe, quantified in
Sec.~\ref{sec:limits}, are consistent with the regimes identified above.

\section{Results}\label{sec:results}

Our system is made up of an aluminum flute cavity and an aluminum transmon and readout resonator evaporated on a sapphire chip. We chose $\Omega_R/2\pi = 6~\mathrm{MHz}$. For the readout resonator, its parameters were $\omega_r/2\pi =7.706~\mathrm{GHz}$, $\kappa/2\pi=0.42~\mathrm{MHz}$ and $\chi_r/2\pi=0.36~\mathrm{MHz}$. For Memory 1 mode $\omega_1/2\pi=6.915~\mathrm{GHz}$, $T_{1_1} = 145~\mu\mathrm{s}$, and $\chi_1/2\pi=35~\mathrm{kHz}$. For Memory 2 mode $\omega_2/2\pi=5.333~\mathrm{GHz}$, $T_{1_{2}} =136~\mu\mathrm{s}$, and $\chi_2/2\pi=20~\mathrm{kHz}$. The parameters of the qubit are $\omega_q/2\pi=6.269~\mathrm{GHz}$, $T_{1_q}=22.8~\mu\mathrm{s}$, and $T_{2_q}^{\mathrm{echo}}=21.8~\mu\mathrm{s}$. The drives were chosen such that $g_1/2\pi=g_2/2\pi=0.182~\mathrm{MHz}$ and for reset pulses $g_r/2\pi=0.456~\mathrm{MHz}$. A full system wiring diagram is shown in Appendix~\ref{app:wiring}.

The main measurement involved Wigner characteristic function measurements of one or both memory modes. The complete measurement process, along with the pulses, is described in Appendix~\ref{app:pulses}.

\subsection{Fock State Generation Measurements}\label{sec:resFockGen}

For each target state $|n\rangle$, $n = 1,\ldots,5$, we scanned the Rabi-drive
duration about the analytically predicted value $\tau_n = \tau_1/\sqrt{n}$
and recorded $\langle \hat{\sigma}_z \rangle$; the optimum
$\tau_n$ corresponds to the extremum of this signal. The scans, shown in
Fig.~\ref{fig:GeneratedFocks}(a), exhibit clear extrema at the predicted
times: extracted values are $\tau_1 = 1.42$, $\tau_2 = 1.01$, $\tau_3 = 0.81$,
$\tau_4 = 0.69$, and $\tau_5 = 0.61~\mu$s, all within a few percent of
$\tau_1/\sqrt{n}$. While we used these calibrated values for the reported Fock-state measurements, the $\langle \hat{\sigma}_z \rangle$ extrema are sufficiently broad that similar performance is expected from the analytically calculated times. This direct experimental confirmation establishes the protocol's defining practical property: once $\tau_1$ is determined, the entire family of higher-$n$ operations is fixed with no further pulse optimization.

Wigner characteristic function measurements of the generated states are shown
in Fig.~\ref{fig:GeneratedFocks}(b). Maximum-likelihood reconstruction yields
state fidelities of $91.64\%$, $82.38\%$, $75.82\%$, $69.39\%$, and $62.98\%$ for
$|n{=}1\rangle$ through $|n{=}5\rangle$, respectively. Because the protocol
employs constant-amplitude, analytically-timed pulses, these fidelities serve
as a direct probe of the hardware's coherence limits during the interaction
rather than of control complexity. As we quantify via full master-equation
simulation in Sec.~\ref{sec:limits}, most of the infidelity is consistent with
the expected ramp-induced dephasing and $T_2$ limits of the present device,
confirming that the analytical scaling performs as expected with no hidden
control errors.

\subsection{Fock State SWAP Measurements}\label{sec:resFST}

We characterize the inter-mode SWAP on three classes of input, in
increasing order of demand on the gate: the vacuum state, the
single-photon Fock state, and small coherent states whose transfer
probes the gate's action beyond the computational-basis populations.

We first optimized the SWAP time $\tau'_1$ via $\langle\hat{\sigma}_z\rangle$
scans, yielding $\tau'_1 = 2.27~\mu$s, in agreement with the theoretical
$\tau'_1 = \sqrt{2}\,\tau_1$ (Fig.~\ref{fig:FST}(a)).

For a vacuum input, both memory modes are initialized in $|0\rangle$ and
the SWAP is applied. The reconstructed states of Memory~1 and Memory~2
following the operation (Fig.~\ref{fig:FST}(b)) remain in vacuum with
fidelities of $96.10\%$ and $97.62\%$, respectively, confirming that the
SWAP does not spuriously populate either mode.

For a single-photon input, Memory~1 is prepared in $|1\rangle$
(reconstructed fidelity $F_{\mathrm{in}} = 91.64\%$) and swapped into the
vacuum of Memory~2. The reconstructed state of Memory~2 following the
SWAP (Fig.~\ref{fig:FST}(c)) has target-state fidelity
$F_{\mathrm{out}} = 68.44\%$. Because the input is itself imperfectly
prepared, $F_{\mathrm{out}}$ is the product of preparation and SWAP
errors. Under the assumption that these
infidelities act multiplicatively and independently, the SWAP step alone
contributes an effective state-transfer fidelity of
$F_{\mathrm{out}}/F_{\mathrm{in}} \approx 74.68\%$ on this input. This
single-input number is not an average gate fidelity in the standard
sense, but together with the vacuum result it constrains the action of
the gate on the computational basis $\{|0\rangle, |1\rangle\}$.

Finally, we demonstrate that the SWAP acts on the single-excitation component
of an arbitrary input state, not only on a deliberately prepared Fock state. We
initialized Memory~1 in small coherent states ($|\alpha| = 0.50$ and
$|\alpha| = 0.77$), dominated by their $|0\rangle$ and $|1\rangle$ components,
and tracked the transfer of the $|1\rangle$ population through the gate
(Fig.~\ref{fig:SwapCoherent}). The $|1\rangle$-component fidelity is preserved
across the operation ($20.34\% \to 15.37\%$ and $31.76\% \to 28.76\%$), the
small absolute values simply reflecting that $|1\rangle$ is by construction a
minority component of these weak displacements. As expected for a gate timed to
the single-photon transfer at $\tau'_1$, the higher Fock components
$|n{\geq}2\rangle$ present in a coherent input evolve with distinct,
$n$-dependent SWAP times and phases (Sec.~\ref{sec:theoFST}), so the output is
not a simply displaced state and we quote $|1\rangle$-population survival rather
than a global phase. The full coherence of the gate within the single-excitation
$\{|0\rangle,|1\rangle\}$ manifold relevant to dual-rail encoding is established
directly by the Bell-state experiment below.

The Bell-state experiment provides a direct, two-mode test of coherence
preservation under the SWAP. We generated the dual-rail Bell state by the
process of Sec.~\ref{sec:theoFST} and measured its joint Wigner
characteristic function via simultaneous conditional displacements on both
memory modes (Fig.~\ref{fig:JointChar}(a)). We continued the sideband drives to apply SWAP and
re-measured (Fig.~\ref{fig:JointChar}(b)). Both agree with the theoretical
prediction (Fig.~\ref{fig:JointChar}(c)) up to measurement noise and
ramp-induced dephasing of the kind described in Sec.~\ref{sec:limits}.
The off-diagonal correlations visible in the $\mathrm{Re}(\alpha)$--$\mathrm{Re}(\beta)$
and $\mathrm{Im}(\alpha)$--$\mathrm{Im}(\beta)$ projections demonstrate
coherence between the two memory modes. The persistence of these
correlations after the SWAP (Fig.~\ref{fig:JointChar}(b)) is the clearest
evidence in this work that the SWAP preserves inter-mode coherence: the
entangled state survives the operation rather than being projected or
dephased.

\begin{figure*}
    \centering
    \includegraphics[width=1\linewidth]{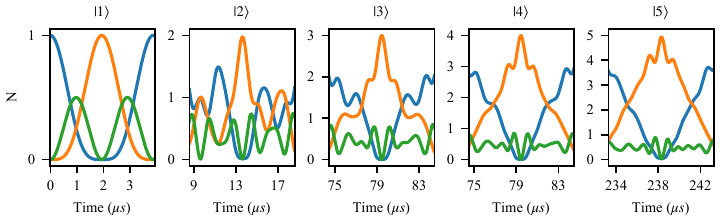}
    \caption{Simulations of the number of excitations as a function of time during Fock state transfer for different initial Fock states. The Hamiltonian used is the one in Eq.~\ref{eq:SimpleHamilt} with $g_r=0$. Memory 1, Memory 2, and the qubit are shown in blue, orange, and green, respectively.}
    \label{fig:FSTsim}
\end{figure*}

\begin{figure*}
    \centering
    \includegraphics[width=0.95\linewidth, trim = 10 20 25 0, clip]{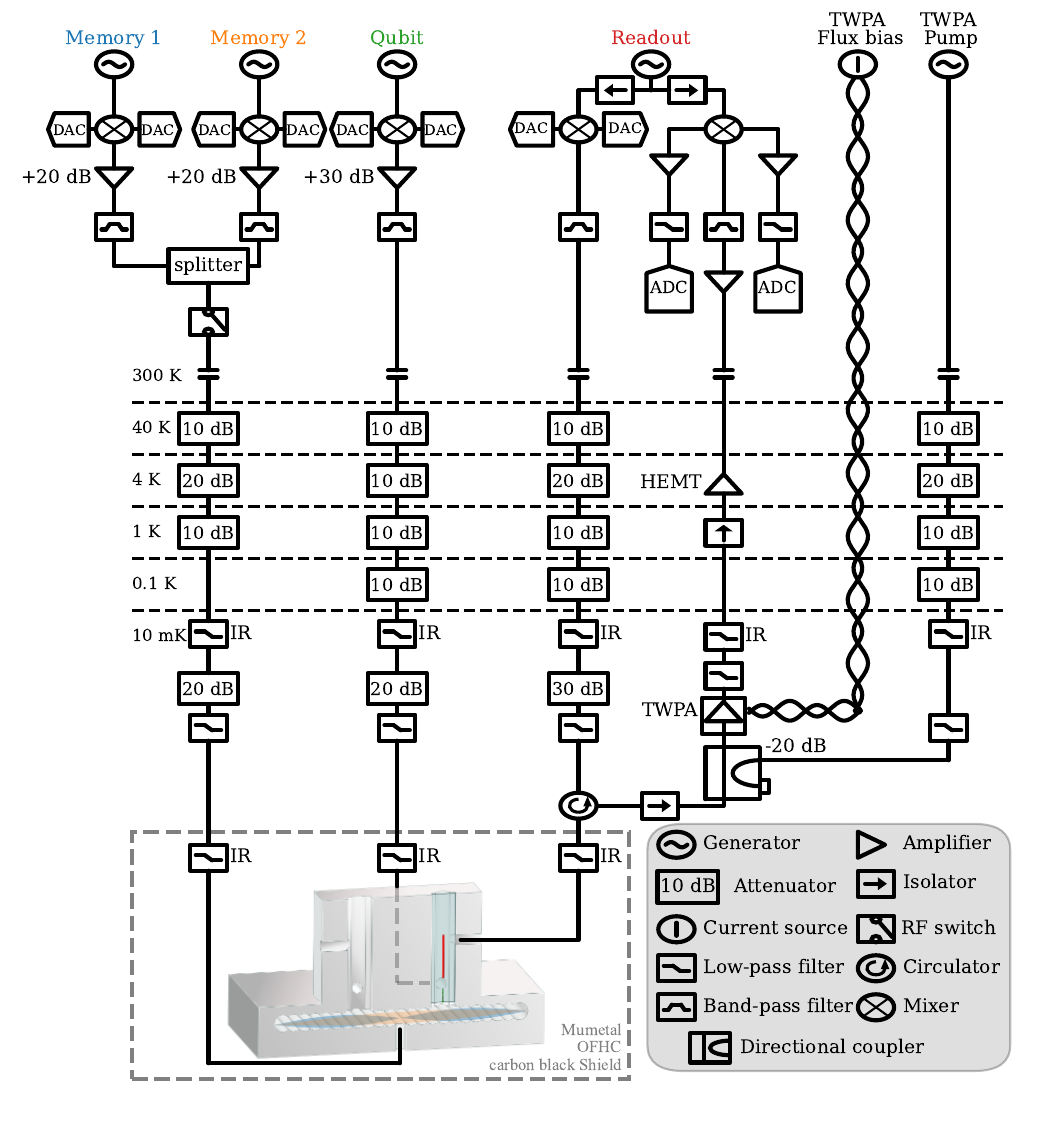}
    \caption{System Wiring Schematics}
    \label{fig:Wiring}
\end{figure*}

\begin{figure}
    \centering
    \includegraphics[width=1\linewidth]{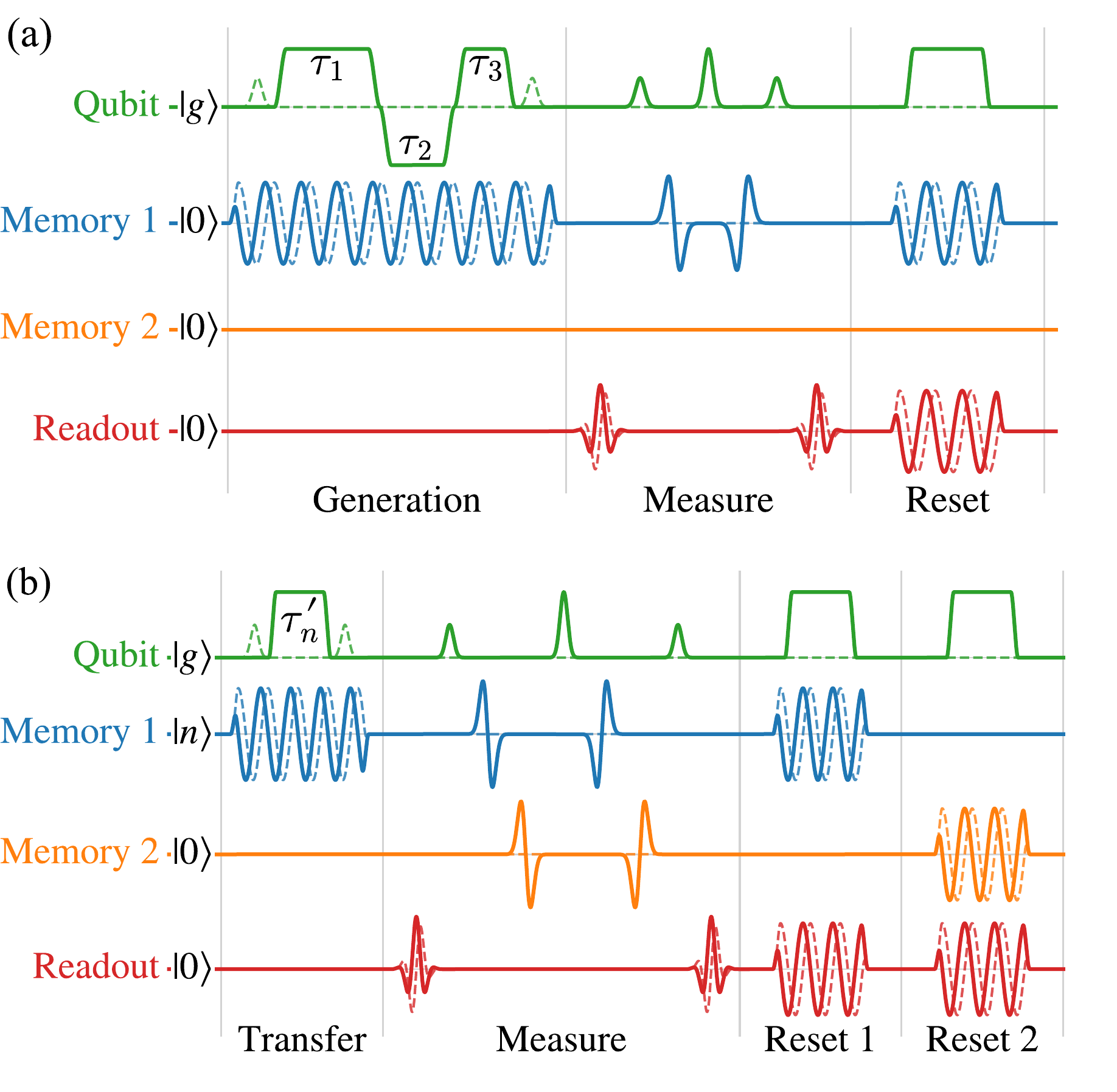}
    \caption{Schematic pulse diagram. Measurement refers to the measurement of the Wigner characteristic function using a conditional displacement as in~\cite{touzard_stabilization_2019}. Conditional displacement in this work was realized using an asymmetric pulse similar to~\cite{eickbusch_fast_2022}. Reset was performed using the method in~\cite{karaev_cavity-mode_2025,blumenthal_experimental_2026}. Solid and dashed lines represent in-phase and dashed signal components, respectively. (a) Fock state generation of $\left|n=3\right>$. (b) Fock state SWAP of state $\left|n\right>$.}    \label{fig:pulses}
\end{figure}

\section{Limits to Fidelity and Projected Performance}\label{sec:limits}

Having validated the analytical scaling law $\tau_n = \tau_1/\sqrt{n}$ experimentally, we now quantify the hardware requirements necessary for this protocol to reach state-of-the-art fidelities. We performed master-equation
simulations of the full Fock-state generation sequence under four
parameter sets that isolate the contributions of ancilla coherence and
of drive ramp times. We then identify the dominant infidelity factors
and quantify the headroom available within the protocol itself.

\subsection{Master-equation simulation of the generation protocol}

We simulate the drive-frame Hamiltonian of Eq.~\ref{eq:SimpleHamilt}
with standard Lindblad collapse operators describing qubit relaxation,
qubit dephasing, and cavity-mode decay. The experimental ramps are simulated as well. We chose $200~\mathrm{ns}$ ramps in the experiment as
a compromise: short enough that the operation completes within the
ancilla coherence budget, but long enough to avoid spectral leakage
into nearby transitions and TLS-rich regions of our device's spectrum.

To isolate the contribution of each effect, we simulated four parameter
sets, shown in
Fig.~\ref{fig:simFockGen}: (a) our device parameters with $200~\mathrm{ns}$
ramps, reproducing the experimental conditions; (b) our device
parameters with $20~\mathrm{ns}$ ramps, isolating the contribution of
ramp-induced dephasing; (c) $10\times$-improved ancilla and cavity
coherence with $200~\mathrm{ns}$ ramps, isolating the contribution of
decoherence; and (d) both improvements combined. The chosen
$10\times$ scaling brings the parameters into a regime routinely
demonstrated in current 3D circuit-QED experiments
($T_{1}^{q} \sim 200~\mu\mathrm{s}$, cavity $T_{1} \sim 1.5~\mathrm{ms}$),
and the $20~\mathrm{ns}$ ramps are typical for devices with spectrally
uncluttered landscapes.

Three observations follow from the simulation. First, the predicted
$|n{=}5\rangle$ fidelity at our device parameters ($74.46\%$) closely
tracks the measured value ($62.98\%$); the residual gap of order $10\%$
is consistent with parasitic couplings to TLSs and higher-order transmon
transitions during the strong sideband drives, couplings that are
known to be present in our spectrally crowded device but are not
included in the master-equation model. This agreement validates that
the dominant infidelity sources in the experiment are already captured
by the included physics, with no need to invoke a methodological limit
of the protocol. Second, shortening the ramps alone raises the predicted $|n{=}5\rangle$ fidelity
from $74.46\%$ to $84.11\%$; improving ancilla coherence alone yields a comparable, slightly larger gain to $90.70\%$. The substantial contribution of ramp-induced dephasing is consistent with the observation that the discrete fidelity
drops in Fig.~\ref{fig:GeneratedFocks}(b) are larger than would be
explained by the $\tau_n = \tau_1/\sqrt{n}$ time increments alone. We
note that in principle a virtual rotation tracking the time-dependent
phase could compensate this dephasing; in our device we were
unable to characterize the relevant phase trajectory with sufficient
accuracy to implement this. Third, the combination of improved
coherence and shorter ramps brings $|n{=}5\rangle$ fidelity to $96.90\%$,
indicating that the protocol itself does not impose any obstacle to
high-$n$ Fock-state generation: the limit is set by the device, not
the methodology.

\subsection{Identified infidelity factors}

We now summarize the three identifiable factors limiting the
experimental fidelities, in decreasing order of impact.

Ancilla coherence during the interaction: The protocol
routes all excitations through the transmon's $g$--$e$ manifold. The
total time spent in this manifold scales as
$\tau_1 \sum_{k=1}^{n} 1/\sqrt{k}$, which for $n{=}5$ is approximately
$3.5\,\tau_1 \sim 5~\mu\mathrm{s}$: a substantial fraction of
$T_{2_q}^{\mathrm{echo}} = 21.8~\mu\mathrm{s}$ in our device. The protocol
is, at our chosen drive power, most naturally implemented on devices with
$T_2^{q} \gtrsim 100~\mu\mathrm{s}$; the simulation predicts fidelity
gains of order $16\%$ at $n{=}5$ from this factor alone (panel (c)
of Fig.~\ref{fig:simFockGen}), making it the single largest contributor
to current infidelity. Alternatively, on systems specifically designed to receive high power drives, it would be possible to perform our protocol much more quickly, leading to a similar increase in fidelity.

Drive ramp times and ramp-induced dephasing: The
time-dependent phase accumulation during the $200~\mathrm{ns}$ drive ramps is
a substantial contributor to infidelity, second only to ancilla
decoherence: shortening the ramps to $20~\mathrm{ns}$ alone raises the
predicted $|n{=}5\rangle$ fidelity from $74.46\%$ to $84.11\%$. The
constraint on ramp duration in our device originates in spectral
crowding rather than in the protocol itself; in devices with cleaner
spectral regions this factor can be substantially mitigated.

Parasitic couplings: The residual $\sim\!10\%$ gap between
the simulation at our parameters (panel (a) of
Fig.~\ref{fig:simFockGen}) and the measured fidelity is consistent
with parasitic couplings to TLSs and higher-order transmon transitions
during the strong sideband drives. These effects are not included in
the bare-decoherence simulation and can be reduced by spectral
engineering of the cavity environment and by avoidance of drive
amplitudes that bring multi-photon transitions into resonance.

\subsection{Headroom within the protocol}

The protocol itself imposes only one fundamental constraint: the
rotating-wave approximation requires $g \ll \Omega_R$. With
$\Omega_R/2\pi = 6~\mathrm{MHz}$ and $g/2\pi = 0.182~\mathrm{MHz}$ in
the present device, we operate well within this limit and have
headroom to increase $g$ (and thus shorten $\tau_1$) on devices that
tolerate stronger Rabi drive without exciting parasitic transitions.
Combining the three improvements above with this headroom, the
simulation predicts $|n{=}5\rangle$ fidelities exceeding $95\%$ with
per-photon times below $1~\mu\mathrm{s}$, placing the protocol within
striking distance of state-of-the-art Fock-state demonstrations while
retaining its parameter economy. While the Hilbert space of the
two-mode SWAP simulation is computationally prohibitive to evolve
directly, the underlying $g$--$e$ interaction is shared between the
generation and SWAP protocols, so we expect comparable improvements to
apply to SWAP fidelities in optimized devices.

\section{Conclusion}

We have introduced and experimentally demonstrated an analytical,
calibration-light protocol for Fock-state generation and inter-mode SWAP on
weakly dispersively coupled cavity modes, operating entirely within the
$g$--$e$ manifold of a Rabi-driven transmon. The protocol's distinguishing
feature is methodological: a single closed-form scaling
$\tau_n = \tau_1/\sqrt{n}$ governs the entire family of pulse times, so
addressing Fock states up to $|n{=}5\rangle$ requires no additional pulse
calibration beyond what is needed for $|n{=}1\rangle$. The same mechanism,
applied symmetrically to two modes, realizes an analytically tractable
inter-mode SWAP; halting the same drive at half its duration produces
a maximally entangled dual-rail Bell state.

Because the protocol uses constant-amplitude, analytically-timed pulses with no
per-state optimization, its fidelity is bounded directly by ancilla coherence
and drive-ramp duration rather than by control complexity. Master-equation
simulation confirms this: the measured fidelities ($91.64\%$ at $n{=}1$ to
$62.98\%$ at $n{=}5$, and $68.44\%$ for single-photon SWAP) are reproduced by a
model containing only the device's known decoherence and ramp parameters, and
the same model predicts fidelities exceeding $95\%$ at $n{=}5$ once these are
brought to values already routine in 3D circuit QED. The analytical scaling
itself thus imposes no barrier to high-$n$, high-fidelity operation. We further
characterize the SWAP on vacuum, single-photon, and coherent-state inputs, and
verify preservation of inter-mode coherence within the single-excitation
manifold through the generation and swapping of an entangled dual-rail Bell
state.

More broadly, this protocol occupies a distinct and useful point in the design
space of bosonic control. Where charge-driven $g$--$f$ schemes~\cite{huang_fast_2026} trade
calibration overhead for multimode reach and direct code-state preparation, and
numerical optimal control trades analytical transparency for raw fidelity, the
present approach prioritizes parameter economy and analytical tractability: a
single closed-form scaling law, the $g$--$e$ manifold alone, and one sideband
line per mode. For systems in which calibration cost, ancilla-level
limitations, or analytical pulse design are paramount, this combination is, to
our knowledge, not offered by any existing protocol. We expect these approaches
to remain complementary as the field matures.

\begin{acknowledgments}
This research was supported by the Israeli Science Foundation grant No. 657/23, and Technion's Helen Diller Quantum Center.
\end{acknowledgments}

\section*{Data Availability}

Data supporting the findings of this study are available upon reasonable request.

\section*{Appendices}

\appendix

\section{Fock State Transfer Simulations}\label{app:simFST}

Simulations of Fock state SWAP for $n\leq5$ are presented in Fig.~\ref{fig:FSTsim}. The simulations are performed using the Hamiltonian in Eq.~\ref{eq:SystemHamilt}, with $g_r=0$, as described in Sec.~\ref{sec:theoFST}. Since for $\left|n\geq2\right>$ the dynamics of the Hamiltonian are more complicated than the simple dynamics of $\left|n=1\right>$, the time $\tau'_n$ cannot be trivially estimated, as explained in Sec.~\ref{sec:theoFST}. However, using simulations we can extract an initial value for the transfer time which can then be optimized using $\left<\hat{\sigma}_z\right>$ measurements for different transfer times, as in Sec.~\ref{sec:resFST}. It is important to note that the non-ideality of $\left|n\geq2\right>$ SWAP is shown in the final photon counts which approach but do not fully reach the initial values.

\section{Bright- and Dark-Mode Formalism}\label{app:BDM}
Let us theoretically derive our method for SWAP and for Bell state generation. Let us denote the state vector as $\left|n_1,n_2,q\right>$, where $n_1$, $n_2$, and $q$ refer to the state of Memory 1, Memory 2, and the qubit, respectively. In the simplified Eq.~\ref{eq:SimpleHamilt} frames, for a single photon, we initialize the system in the state

\begin{equation}
    \begin{split}
    \left|\psi\right>_{init}=\left|1,0,g\right>,
    \end{split}
    \label{eq:initBell}
\end{equation}

where $\left|g\right>$ here denotes is the lower-energy dressed state in the lab frame. We explicitly rewrite Eq.~\ref{eq:SimpleHamilt} for $g_1=g_2=g$ and $g_r=0$ by using the bright-mode and dark-mode operators

\begin{equation}
    \begin{split}
        \hat{B}\equiv\left(\hat{a}_1+\hat{a}_2\right)/\sqrt{2},\quad
        \hat{D}\equiv\left(\hat{a}_1-\hat{a}_2\right)/\sqrt{2},
    \end{split}
    \label{eq:BrightDark}
\end{equation}

so that the Hamiltonian becomes

\begin{equation}
    \begin{split}
        H=\sqrt{2}g\left(\hat{B}^\dagger\sigma_-+\hat{B}\sigma_+\right).
    \end{split}
    \label{eq:BellHamilt}
\end{equation}

Let us denote the state vector in the bright and dark mode basis as $\psi=\left|n_B,n_D,q\right>_{BD}$, where $n_B$ and $n_D$ are the number of bright- and dark-mode excitations. The initial state is then expressed as

\begin{equation}
    \begin{split}
    \left|\psi\right>_{init}=\frac{\left|1,0,g\right>_{BD}+\left|0,1,g\right>_{BD}}{\sqrt{2}}.
    \end{split}
    \label{eq:initBellBD}
\end{equation}

The dark mode does not evolve, and the bright mode excitation oscillates between the qubit and the bright mode at a rate $\sqrt{2}g$. Thus the state at time $t$ becomes

\begin{equation}
    \begin{split}
        \left|\psi(t)\right>=  \frac{\mathrm{cos}\left(\sqrt{2}gt\right)\left|1,0,g\right>_{BD}}{\sqrt{2}}\\-\frac{i\mathrm{sin}\left(\sqrt{2}gt\right)\left|0,0,e\right>_{BD}}{\sqrt{2}}\\+\frac{\left|0,1,g\right>_{BD}}{\sqrt{2}}.
    \end{split}
    \label{eq:evolvingBDState}
\end{equation}

The excitation will be transferred to Memory 2 if the initial state's bright-mode excitation gains a phase of $\pi$, i.e. $\hat{B}\rightarrow-\hat{B}$. This would create the state

\begin{equation}
    \begin{split}
    \left|\psi\right>_{\mathrm{SWAP}}=\frac{-\left|1,0,g\right>_{BD}+\left|0,1,g\right>_{BD}}{\sqrt{2}},
    \end{split}
    \label{eq:swappedBD}
\end{equation}

which in the original basis is 

\begin{equation}
    \begin{split}
    \left|\psi\right>_{\mathrm{SWAP}}=-\left|0,1,g\right>.
    \end{split}
    \label{eq:swappedBell}
\end{equation}

From the $\left|\psi(t)\right>$ expression it is possible to see that this occurs at time $\tau'_1\equiv\pi/(\sqrt{2}g)$. Since the original operators are expressed in the bright- and dark-mode formalism as

\begin{equation}
    \begin{split}
        \hat{a}_1=\left(\hat{B}+\hat{D}\right)/\sqrt{2},\quad -\hat{a}_2=\left(-\hat{B}+\hat{D}\right)/\sqrt{2},
    \end{split}
    \label{eq:unBrightDark}
\end{equation}

It is possible to see that each swapped photon adds a factor of $-1$. This shows that for the Hamiltonian in Eq.~\ref{eq:SimpleHamilt} the swapping will be a SWAP gate for even-numbered photon counts and a -SWAP gate for odd-numbered photon counts.

 If we instead initialize both memory modes at vacuum and the qubit in the higher-energy dressed state, such that the initial state is

 \begin{equation}
    \begin{split}
    \left|\psi\right>_{init}=\left|0,0,e\right>_{BD}.
    \end{split}
    \label{eq:initBellBD2}
\end{equation}
 
The state evolves as

\begin{equation}
    \begin{split}
        \left|\psi(t)\right>=  \frac{\mathrm{cos}\left(\sqrt{2}gt\right)\left|0,0,e\right>_{BD}-i\left|1,0,g\right>_{BD}}{\sqrt{2}},
    \end{split}
    \label{eq:evolvingBDStateBell}
\end{equation}

and if we let it evolve for $t=\tau'_1/2=\pi/(2\sqrt{2}g)$, the final state becomes, up to a global phase,

\begin{equation}
    \begin{split}
        \left|\psi\right>=\frac{\left|1,0\right>+\left|0,1\right>}{\sqrt{2}},
    \end{split}
    \label{eq:postBell}
\end{equation}

which is a maximally entangled Bell state.

\section{System Wiring Schematics}\label{app:wiring}

The wiring diagram of the system is shown in Fig.~\ref{fig:Wiring}

\section{Pulse Procedure}\label{app:pulses}

We performed measurements of both Fock state generation and SWAP, including Bell state generation.
The measurements involved measuring the Wigner characteristic function of one or both of the memory modes. The measurement of the Wigner characteristic function is performed by initializing the qubit with a $\pi/2$ pulse, followed by a conditional displacement $\hat{D}(\alpha\hat{\sigma}_z)$ operation. This is followed by a $\pi$ pulse and an additional $\hat{D}(\alpha\hat{\sigma}_z)$ conditional displacement in order to improve the effective dephasing rate. Following these pulses the real and imaginary parts of the Wigner characteristic function are encoded in the non-$z$ axes expectation values of the qubit, and thus an appropriate $\pi/2$ pulse along one of those axes is performed along with a measurement in order to extract the Wigner characteristic function. This is described in more detail in~\cite{touzard_stabilization_2019}.
This operation is followed by a cavity reset for all participating memory modes using the method described in~\cite{karaev_cavity-mode_2025,blumenthal_experimental_2026}.

Additionally, immediately following generation or SWAP, we performed a measurement of the qubit. Since both methods are designed such that the qubit is supposed to be in the $\left|g\right>$ state following the operation, we used this measurement to post-select on this outcome, discarding measurements in which the qubit was instead in the $\left|e\right>$ state.

\newpage
\bibliography{references}

\end{document}